\documentclass[preprint]{JHEP3} % 10pt is ignored!

\newcommand\fverb{\setbox\pippobox=\hbox\bgroup\verb}
\newcommand\fverbdo{\egroup\medskip\noindent%
            \fbox{\unhbox\pippobox}\ }
\newcommand\fverbit{\egroup\item[\fbox{\unhbox\pippobox}]}
\newbox\pippobox                                                                  %

\usepackage{epsfig}
\usepackage{graphicx}
\usepackage{wrapfig}
\usepackage{boxedminipage}
\usepackage{setspace}
\usepackage{subfigure}

 \def\eq#1{(\ref{#1})}
 \def\ap{\alpha'}
\def\d{\partial}

\def\half{\frac{1}{2}}

\def\){\right)}
\def\({\left( }
\def\]{\right] }
\def\[{\left[ }
 
\newcommand{\ddd}{\cdot\cdot\cdot}

\def\half{{1\over 2}}

\newcommand{\be}{\begin{equation}}
\newcommand{\ee}{\end{equation}}
\newcommand{\ba}{\begin{eqnarray}}
\newcommand{\ea}{\end{eqnarray}}

\def\Tr{{\rm Tr}}

\title{Bulk Filling Branes  and the Baryon Density in AdS/QCD with  gravity back-reaction}
\author{Sang-Jin Sin \\
Department of Physics, Hanyang University, 133-791, Seoul, Korea\\
E-mail: \email{sjsin@hanyang.ac.kr }
\thanks{Present address: Department of Physics and Astronomy, SUNY Stony-Brook, NY 11794 }
}

\abstract{
We consider the 
gravity back reaction on the metric due to the baryon density in effective ads/qcd model 
by reconsidering the role of the charged AdS black hole. 
Previously it has been known that the $U(1)$ charge is dual to the R-charge.
Here we point out that if we consider the case where
$AdS_5$ is completely filled with $N_f$ flavor branes, the gravity back reaction produces charged AdS black hole
where the effect of charge on the metric is proportional to  $N_f/N_c$. As a consequence,  phase diagram  changes qualitatively if we allow $N_f/N_c$ finite:
it closes at the finite density unlike the probe brane embedding approach.
 Another issue we discuss here is the question whether
there is any chemical potential dependence in the
confining phase.  We consider this problem in the hard wall model with baryon charge. We conclude that there is a non-trivial dependence on the chemical potential in this case also.}

\begin{document}

\section{Introduction}
It is of great interest to see  how far one can use AdS/CFT\cite{ads/cft,witten} to understand   the strongly interacting nuclear force. So far it has been suggested to  apply  in the strongly interacting  quark gluon plasma\cite{RHIC,son,SZ,Nastase,JP,yaffe}
and phenomenological aspect of chiral dynamics
\cite{KK,PS,evans,myers2,SS,EKSS,Hong}.
Especially interesting is to investigate the baryon density effect \cite{KSZ,HT,NSSY,KMMMT,parnachev,harvey,KLNPS} on the phases of the Yang-Mill theory.

The purpose of the paper is to study the gravity back reaction to the baryon density in  the effective ads/qcd model by reconsidering the role of the charged AdS black hole  in AdS/CFT. Previously this background has been studied\cite{EJM} and its local $U(1)$ charge is identified as dual to the R-charge. See also \cite{cai}. The gravity back reaction for the general flavor branes has been studied in various contexts \cite{liu,erdmenger,kirsch,sonnenschein,nunez1} and usually it is very non-trivial problem.
Here we point out that in the case of effective AdS/QCD model \cite{EKSS}, which can be   
interpreted to have bulk-filling-flavor-branes,  the gravity back reaction is simple and  the result is nothing but the
charged AdS black hole, 
where the effect of baryon charge on the metric is proportional to  $(N_f/N_c) N_B$. 
This $N_f/N_c $ suppression can be attributed to the fact that
R-charge is carried by adjoint scalar while baryon charge is carried by the bi-fundamentals.
If we consider strictly  large $N_c$ limit, the metric correction is negligible. However, if we are interested in $N_f/N_c$ correction, this is important.
Such a $N_f/N_c$  correction to the metric results in corresponding corrections in all physical observables.
Especially, in phenomenological applications
where one calculate using the finite value  of $N_c$  as well as finite  $N_f$, this   $N_f/N_c$ correction  will be essential.
One of the consequence of  back reaction of the metric
is the qualitative change in phase diagram:
it closes at the finite density, while in the probe brane embedding approach \cite{HT} or in the hard wall model  without gravity back reaction \cite{KLNPS} the phase diagrams does not close. The main reason for this change can be traced back to the charge dependence of the temperature.

We also consider an issue which is confusing
in the present literature. We first describe the problem
by summarizing the relevant history of baryon density problem in AdS/QCD:
In \cite{KSZ}, it is suggested  that the baryon/quark chemical potential should be treated as the boundary value of the brane electric potential $A_0$
and electric potential was determined by the vector domination.
In \cite{HT}, it was suggested that one can use equation of motion of DBI action to determine the electric potential. Non-trivial Electric potential can exist only with some sources. However, the authors of \cite{HT} imposed  a smoothness  condition on the electric potential $A_0$ for the in the chiraly asymmetric phase. As a result, constant potential was obtained. This is the same as assuming that there is no source on the brane.
In \cite{parnachev}, the same boundary condition is chosen
and it is related to the earlier work on Hawing-Page transition of charged AdS black hole \cite{EJM},
where for the fixed chemical potential, low temperature phase is associated with the  thermal AdS space.
As a consequence,
electric potential is constant and
there is no chemical potential dependence of thermodynamics and average baryon density is zero in that phase phase. We believe this is not physically acceptable.
\footnote{There is another issue regarding baryon density in AdS/CFT, which we will not be treated in this paper and will be mentioned in the discussion section.}

However, the Hawking-Page transition in \cite{EJM} is for
the case with $S^3$ boundary, and for the flat boundary, there is no phase transition. Therefore there is no necessity to consider thermal AdS for the low temperature, and the work in \cite{EJM} is not relevant to this case.
However, when we consider the round boundary or introduce the confining phase with hard wall, this is still an issue. One of the purpose of this paper is to consider above issue in the context of hard wall model.

The rest of the paper is in following order.
In section 2, we set up the AdS/QCD from brane embedding point of view and derive the gauge coupling on the in terms of basic variables $N_c, N_f$ and describe  charged ads black hole as a back reaction of gravity to the flavor branes. We also and
determine the metric dependence of the baryon charge density. In section 3, we describe the thermodynamics of the theory without confinement. In section 4, we discuss the  issue of chemical potential dependence of grand potential and  we explicitly workout the phase diagram in  temperature and chemical potential $(T,\mu)$ for the theory with confinement.
In section, we summarize and describe some future projects.

\section{Baryon density and the gravity back-reaction}
In this section we set up the AdS/QCD from brane embedding point of view and derive the gauge coupling on the in terms of basic variables $N_c, N_f$ and consider  the
metric back reaction to the presence of baryon charge density.
\subsection{Bulk Filling Branes and AdS/QCD. }
We start from the  DBI action of the D7-brane. For a given background and embedding, we can write the DBI action interms of the induced metric $g_{ab}$ and gauge field on the brane.
\footnote{The Chern-Simons term vanishes in the
present case.}
\begin{eqnarray}
I_{DBI}=-N_{f}\mu_{7}
\int d\sigma^8 \sqrt{-\det(g_{ab}+2\pi \alpha' F_{ab})}, \label{action}
\end{eqnarray}
where $\mu_{7}$ is the D7-brane tension
 The   D7 embedding of our concern  is those wrapping $S^3$ of $S^5$
 and cover our own spacetime. If we neglect the $S^3$ dependence of the field variable, \eq{action} defines a 5 dimensional model.
The DBI action can be expanded in the power of $\alpha' F_{ab}$ and the quadratic part is
\ba
I_{2}=-N_{f}\mu_{7}(2\pi \alpha')^2  \Omega_3 \cdot\frac{1}{4}
\int d^5\sigma \sqrt{-\det g^{(8)}} F_{ab}F^{ab},\label{quad}
\ea
 where  $\Omega_3$ is the volume of $S^3$ and we have used
$
\sqrt{\det(1+A)}=1-\frac{1}{4} \Tr A^2 -\frac{1}{8}\Tr{A^4}+\frac{1}{32}(\Tr A^2 )^2+\ddd,$
 for traceless matrix $A=g^{-1}F$.
 Eq. \eq{quad} can be used to define phenomenological models as an approximation to the original DBI action.

We first take the D3 background in the near horizon limit:
\ba
ds^{2}
&=&\frac{r ^{2}}{R^{2}}\left(-  dt^{2}+d\vec{x}^{2}\right)
+
R^{2}\left(\frac{dr ^{2}}{  r ^{2}}+d\Omega_{5}^{2}\right),\\
&=&\frac{r^{2}}{R^{2}}\left(-dt^{2}+d\vec{x}^{2}\right)
+
\frac{R^{2}}{r^{2}}\(
 d\rho^{2}+\rho^{2}d\Omega_{3}^{2}+dy^{2}
+y^{2}d\varphi^2 \),\nonumber
\ea
In this background, the flat embedding $y=y_0=2\pi \alpha' m_q$ is allowed, for which $r^2=y_0^2+ \rho^2$ and the induced metric on the brane is given by
\begin{eqnarray}
ds_{D7}^{2}
=
\frac{r^{2}}{R^{2}}\left(-dt^{2}+d\vec{x}^{2}\right)
+
\frac{R^{2}}{r^2}
\left(
d\rho^{2}+\rho^{2}d\Omega_{3}^{2}\right). \label{indm1}
\end{eqnarray}
For this case the volume factor is simplified
$\sqrt{-\det g^{(8)} }=\rho^3$.
One should notice that if we further assume that
$y_0=0$,  then the D7 fills all the $AdS_5$ and  the background radial coordinate $r$ is identical to the world volume radial coordinate,  the induced metric.
is identical to the AdS bulk metric.
The action now  becomes precisely the gauge action part of AdS/QCD model of \cite{EKSS}
 \be
I_{guage}=-\frac{1}{4g^2} \int d^{4}x dr
 \sqrt{-\det g^{(5)}} F_{\mu\nu}F^{\mu\nu}\label{gauge1}
 \ee
with identification
\be
\frac{1}{g^2}=\mu_7 \Omega_3 (2\pi \alpha')^2R^3=\frac{N_cN_f}{(2\pi)^2 R},
\ee
 and
$  \quad \sqrt{-\det g^{(5)}}=\sqrt{-\det g^{(8)}}/R^3.$
Notice that the action contains the metric through the
volume factor as well as through two $g^{\mu\nu}$ factors in the $F^{\mu\nu}$. For a general background and a general embedding, Eq.\eq{gauge1} will contain the correction coming from the difference  of induced metric from the background metric. One  can  build models considering such corrections, which is a deviation of
original hard wall model.

Here,  instead of considering such correction we ask following question:  When the probe brane is
filling the 5 dimensional bulk which is asymptotically AdS space, can the induced metric be the same with the bulk metric?  The answer positive.
For AdS Schwarzschild case, we can work out the answer explicitly.
Take the black brane solution in the form where
 the radial coordinate of bulk and brane is the same and also D7's winding structure is manifest:
\ba
ds^{2}
&=&\frac{r ^{2}}{R^{2}}\left(- f(r) dt^{2}+d\vec{x}^{2}\right)
+
R^{2}\(\frac{dr ^{2}}{ f(r) r ^{2}}+
d\theta^{2}+\sin^2\theta d\psi^2+\cos^2\theta d\Omega_{3}^{2}  \),
\ea
Then the induced metric is specified if we determine $\theta$ in terms of  $r$:
\ba
ds^{2}_{D7}
&=&\frac{r ^{2}}{R^{2}}\left(- f(r) dt^{2}+d\vec{x}^{2}\right)
+
R^{2}\( \(\frac{1}{ f(r) r ^{2}}+
 {\theta'(r)}^{2}\)dr ^{2}+ \cos^2\theta d\Omega_{3}^{2}  \),
\ea
For the space filling embedding we are interested,
\be r\sin\theta=2\pi\ap m_q \to 0,\ee
  therefore  $\theta'=0 $.
In this case, the  world volume measure factor is reduced to that of 5 dimensional background metric:
\be
\sqrt{-\det g^{(8)}}/R^3\to \( {r}/{R}\)^3
\ee
which is precisely the volume factor of $AdS_5$ Schwarzschild  solution.

For more practical purpose,  we can simply assume that \eq{gauge1} is the
definition of the model and add  necessary terms like scalars and without further consideration of brane embedding \cite{EKSS}. The rest of the paper is essentially following this philosophy by taking the space filling branes
as an assumption.

\subsection{The baryon  chemical potential and charged AdS black hole.}
The chemical potential can be treated as the tail of the electric potential living on the probe brane\cite{KSZ,HT}.
In the brane embedding approach, considering  back reaction of the metric  is very hard to study  analytically.
 However,  when the brane is completely filling the background space-time, it can be easily done.
This is because there is a unique way  for the $U(1)$ gauge potential to couple to the gravity: the solution is nothing but the Reisner-Nordstrom metric considered in ref. \cite{EJM}, where the charge was identified as the R-charge, whose carrier is adjoint representation of $N=4$ gauge theory.   Therefore its gravitational coupling's order of magnitude goes like $1/8\pi G \sim N_c^2  $.
In our case the charge is weighted by   $N_f/N_c$, and so is the back reaction.

Our starting point is to add gravity plus
$U(1)$ gauge interaction term to the
phenomenological model of AdS/QCD \cite{EKSS} action:
\be
I=\frac{1}{2\kappa^2}\int d^5x\sqrt{-g}\( {\cal R}+\frac{12}{R^2} \)- \frac{1}{4g^2}\int d^5x\sqrt{-g}F_{\mu\nu}F^{\mu\nu} + I_{flavor}[A_L,A_R;\Phi].
\ee
with $\kappa^2=8\pi G_5$, and then  neglect the original scalar and non-abelian flavour gauge interaction part
$I_{flavor}[A_L,A_R;\Phi]$. Since we want to utilize the known charged AdS balck hole solution \cite{EJM}, the ansatz for the metric and the potential is
\ba
ds^{2}
&=&\frac{r ^{2}}{R^{2}}\left(- f(r) dt^{2}+d\vec{x}^{2}\right)
+
\frac{R^{2}}{r ^{2}} \frac{dr ^{2}}{ f(r) }   ,\quad {\rm with}\quad
f(r)=\frac{r^2}{R^2}-\frac{m}{r^2}+\frac{q^2}{r^4}\\
 A_0&=&\mu-\frac{Q}{r^2}.
 \ea
Notice that $q$ is a parameter describing the metric deformation  while $Q$ is the actual charge.
 We ask what is the precise relation of the parameter $q$ and $Q$ to satisfy the equation of motion:
\ba
{\cal R}_{\mu\nu}-\half g_{\mu\nu} {\cal R} -\frac{6}{R^2} &=&
 {\kappa^2}T_{\mu\nu},\\
T_{\mu\nu}=\frac{-2}{\sqrt{-g}}\frac{\d I}{\d g^{\mu\nu}}&=&\frac{1}{g^2}\( F_{\mu a} F^{\nu a}-\frac{1}{4}g_{\mu\nu}F_{ab}F^{ab}\)+ T_{\mu\nu}^{L,R}.
\ea
Since we consider back reaction to baryon charge, which is the diagonal $U(1)$ charge of the flavor brane, we set non-abelian fields zero: $ T_{\mu\nu}^{L,R}=0.$
The answer is
\be
q^2=\frac{2\kappa^2}{3g^2} {Q^2} = \frac{2}{3}\frac{ N_f}{N_c} R^2 {Q^2}:=a^2R^2Q^2,
\ee
confirming  $N_c/N_f$ dependence as we stated above.
Here we have used  $\frac{R^3}{\kappa^2}=\frac{N_c^2}{4\pi^2}$, and $\frac{R}{g^2}=\frac{N_c N_f}{4\pi^2}$.
This suppression in gravity coupling of the
baryons  relative to  R-charges is, of course,  due to the difference in degrees of freedom between the fundamentals and adjoints. If we consider strictly  large $N_c$ limit, the metric correction is negligible. However, if we are interested in $N_f/N_c$ correction, this is important.
Such a $N_f/N_c$  correction to the metric results in corresponding corrections in all physical observables.
Especially, in phenomenological applications
where one calculate using the finite value  of $N_c$  as well as finite  $N_f$, this   $N_f/N_c$ correction  will be essential and this will most important implication of this work. In this way, the baryon density problem in AdS/QCD can have a simple  description including the back reaction of the metric.
Without metric correction, it is difficult for the AdS/QCD defined as a quadratic action of gauge fields to encode the density effect unless one add
higher order terms (${\cal O}(F^4)$) \cite{KSZ}  or Wess-Zumino term \cite{harvey}.
\vskip 1cm

\section{Thermodynamics without hard-wall.}
For the spherical boundary, there is a Hawking-Page transition (HPT) associated with deconfinement phase transition \cite{witten} and the HPT for the  charged black hole in AdS space (AdSRN) was discussed in \cite{EJM}. For the flat  boundary we are interested here,
there is no other scale than the temperature and therefore there is no such transition: the system is always in a de-confined phase, if we do not install hard wall.
There will be some differences in fixed charge case compared with ref. \cite{EJM} apart from the topology of the boundary.

\subsection{Fixed chemical potential}
The total action is evaluated to  be\footnote{Here we should work with Euclidean action which is obtained by taking a overall - sign in the Minkowski action. To make the electric potential real, we also have to change $F^2\to -F^2$ .}
\be
I =\frac{V_3}{2\kappa^2} \int_{r_+}^\infty dr \int_0^\beta\sqrt{g^{(5)}}\[ \frac{2q^2}{r^6}+\frac{8}{R^2}-\frac{4\kappa^2}{g^2}\frac{Q^2}{r^6}\] . \ee
%\be
%+
%\frac{1}{4g^2}\int d^5x F^2=
%\int\[\frac{2q^2}{r^6}+\frac{8}{R^2} - %\frac{1}{4g^2}\frac{6q^2}{r^6} \]
%.
%\ee
where  $L$ is a large cutoff at large radius.
The last term can be written $ - {6q^2 }/{r^6}$ using
$\frac{Q^2}{g^2}= \frac{3q^2 }{2\kappa^2}.$
If we define the horizon radius $r_+$ as  the larger zero of $f(r)=0$,
we can write the mass in terms of $r_+$
\be
m=\frac{r_+^4}{R^2} +\frac{q^2}{r_+^2}.\label{mass}
\ee
The temperature is defined by the singularity free condition:
\be
T=\frac{f'(r_+)}{4\pi}=\frac{r_+}{\pi R^2} - \frac{q^2}{2\pi r_+^5} = \frac{r_+}{\pi R^2}-\frac{a^2 R^2 \mu^2}{2\pi r_+}.\label{temp}
\ee
where we used  the relation between the chemical potential and the
charge $\mu=Q/r_+^2$, determined by the condition that electric potential is vanishing at the horizon \cite{HR,EJM,HT,NSSY}:
\be
A_0(r_+)=\mu-Q/r_+^2=0,
\ee
and
\be
q=aRQ=aRr_+^2\mu, \quad {\rm with} \quad a=\sqrt{\frac{2N_f}{3N_c}}.
\ee
The action is evaluated to be,
\be
I_{RN} =\frac{\beta V_3}{2\kappa^2R^3}\[ -\frac{q^2}{r^2 }+\frac{2r^4}{R^2}   \]_{r_+}^L,
\ee
which diverge as $L\to \infty$.
We regularize by the value  of the thermal AdS action:
 \be
I_{AdS}=\frac{V_3\beta'}{2\kappa^2R^3}    \frac{2L^4}{R^2}  .
\ee
We determine $\beta' (\simeq \beta) $ such that  the volume of the two space outside cutoff $r=L$ is  equal:
\be
\beta'  =\beta \(1- {mR^2}/{2L^4}\).
\label{match}
\ee
Then,  the regularized action is given by,
\be
I^{(reg)}_{RN}=-\frac{\beta V_3}{2\kappa^2 R^3} \(\frac{r_+^4}{R^2}+\frac{q^2}{r_+^2}\) ,
\ee
where we used \eq{mass}.

Now, we can calculate thermodynamic functions like pressure $P$, energy  $E$ , total Mass $M$, entropy $S$ and total charge $Q_{tot}$ can be calculated to be
\ba
E&=&\(\frac{\d I}{\d \beta}\)_\mu-\frac{\mu}{\beta}\(\frac{\d I}{\d \beta}\)_\beta= 3mb=M=3P,\quad \\
S&=&\beta\(\frac{\d I}{\d \beta}\)_\mu-I= 4\pi r_+^3 b=\frac{A}{4G},\quad \\
\langle Q\rangle &=&-\beta^{-1}\(\frac{\d I}{\d \mu}\)_\beta=6qb,
\ea
where $A$ is volume of horizon and   constant $b$ is given by
\be
 b=\frac{V_3}{2\kappa^2 R^3}=\frac{N_c^2V_3}{8\pi^2 R^6}.
\ee
It is interesting to find an expression of entropy and  energy in terms of temperature and  chemical potential in small charge and high temperature region.
 \ba
s=\frac{S}{V_3}&=&\frac{\pi^2}{2}{N_c^2}{T^3}+\half N_cN_f\mu^2T+\frac{1}{54\pi^4}\frac{N_f^3}{N_c}\frac{\mu^6}{T^3} -\frac{1}{54\pi^6}\frac{N_f^4}{N_c^2}\frac{\mu^8}{T^5}
\cdots,\\
\epsilon=\frac{E}{V_3} &=&\frac{3\pi^2}{8}{N_c^2}{T^4}+ \frac{3}{4} N_cN_f\mu^2T^2+ \frac{N_f^2}{4\pi^2} \mu^4+
\frac{1}{36\pi^4}\frac{N_f^3}{N_c}\frac{\mu^6}{T^2} +\frac{1}{72\pi^6}\frac{N_f^4}{N_c^2}\frac{\mu^8}{T^4}
 \cdots.
\ea
Notice that  coupling dependence is hidden in $N_f$ in  this expression through $R/g^2=N_cN_f/4\pi^2$.
It would be interesting to compare this with the
gauge theory calculation that can be done  along the line of recent paper by Yaffe et.al\cite{yaffe2}.
\subsection{Fixed charge}
For the canonical ensemble,
 The quark number is  conserved under the deconfinement-phase- transition.
Let's assume that $\mu$ as the quark number and treat it as a conserved quantum number under the deconfinement phase transition. We need to add a boundary term \cite{HR} to guarantee the equation of motion with the  fixed  charge.
\be
I_A= -\frac{1}{4g^2}\int d^5x\sqrt{-g} F^2 \to I_A +\frac{1}{g^2}\int_\Sigma F_{\mu\nu}n^\mu A^\nu.
\ee
The added surface term has the  effect that is precisely changing the sign of the electromagnetic action:\footnote{This addition of surface charge does not have essential  effect in discussing the issues above. }
$
I_A+I_{surface}= -I_A
$ at on-shell.
Therefore the value of action $I_{RN}$ with cut off at large radius $L$  is
\ba
{\tilde I}^{(reg)}_{RN}&=&\frac{\beta V_3}{2\kappa^2 R^3}\(  -\frac{ r_+^4}{R^2}+\frac{5q^2}{r_+^2}\) .\label{fixedQRN}\ea
One should notice that the result  is precisely equal to that of Legendre transformation:
\be
{\tilde I}=I-\mu\(\frac{\d I}{\d\mu}\)_\beta.
\ee
This relation is not accidental:  ${\tilde I}$ is describing the fixed charge sector where Gauss law constrains the field configurations and equations of motion should be derived not from the action but its Legendre transformation.
Identifying grand potential $\Omega$ by $I=\beta \Omega$,
the above relation corresponds to the thermodynamic relation  between the free energy $F={\tilde I}/\beta$ and the grand potential $\Omega$.
\be
F=\Omega+\mu {\bar Q}.
\ee
Once we realize the thermodynamic potential in terms of action and its Legendre transformation, we can use them to calculate physical quantities.
\ba
S&=&-\(\frac{\d F}{\d T}\)_q=b\cdot 4\pi r_+^3=\frac{A}{4G},\\
E&=&F+TS=3bm=M,\\
\mu&=&\(\frac{\d F}{\d  Q_{tot}}\)_q=\frac{Q}{r_+^2}.
\ea
It turns out that   canonical (fixed charge) and grand canonical (fixed chemical potential) give completely equivalent description in this case.
In this section, there is no phase transition and  charge free AdS space is used just for  the regularization.
Since the relation between the free energy and grand canonical potential is consequence of charge conservation and gauss law, it should hold regardless of presence of
phase transition.

Our result in this subsection should be compared with the result of ref. \cite{EJM}, where the authors used  extremal charged black hole as a background to subtract, making some difference with the result here. Since extremal black hole can not have arbitrary temperature, it can not be used as a regularizing background without introducing conical singularity.

\section{Charged background with Hard wall: Hawking-Page transition}
To discuss the confinement in the context of D3 background, one has to introduce the IR cutoff, called hard wall \cite{PS,EKSS}.
In gravity background and brane embedding picture, this corresponds to the situation where D7's (as well as the baryon vertices) are expelled by the repulsive core of the confining background\cite{CM}.
The repulsive hard core is singular \cite{CM}, and
physically relevant region must be somewhat away from the singularity because the singularity invalidate the solution's reliability. One has to put a boundary by hand to cut off the unreliable region around the singularity.
The IR brane  in phenomenological  model with  can be understood as this boundary with  background simplified to $AdS_5$.
We assume that probe flavor brane is occupying all the relevant region of bulk of $AdS_5$.

In the interesting paper \cite{herzog},
it was pointed out that in the presence of  IR brane,
HPT exists even for the flat boundary case
and worked out explicitly for the hard wall model without charge \cite{PS,EKSS}.
Here we study the analogue of it in the presence of baryon charges.

\subsection{Fixed charge}

For high temperature, the system is described by a charged
AdS black hole and for low temperature, confining configuration takes over by set-up.
The key question is what is the low temperature pair of charged black hole  for the purpose of the HPT.
There are a few possibilities.

The fist possibility is the to take the extremal AdS RN black hole to take care of the charge conservation. In ref. \cite{EJM}, phase diagram was worked out with this choice. Then
$f'(r_+)=0$ and $f(r_+)$=0 so that for the given value of charge the horizon and the mass  has fixed value
$ m_*=\frac{3}{2}\frac{q^2}{r_*^2},$ and $r_*=2^{-1/6}q^{1/3}R^{1/3}.$
Therefore
\be
I_{extremal~RN}=\frac{\beta_* V_3}{2\kappa^2 R^3}\( \frac{2(L^4-r_*^4)}{R^2}+\frac{4 q^2}{r_*^2}   \).\label{extremalS} \ee
Under this choice,  we need  to take the difference between \eq{fixedQRN} and \eq{extremalS} with a ``common" inverse temperature $\beta$, instead of $\beta_*$ in \eq{extremalS}.
However, while thermal AdS can be in any
temperature, the extremal  black hole  has a fixed temperature, therefore it can not be used as a background at an arbitrary temperature competing  with non-extremal black hole.\footnote{In the round boundary case \cite{EJM} with high chemical potential,  there is no phase transition in high high chemical potential. In the flat boundary limit only high  chemical potential case survive and  the result is reduced to the previous section.
So, although the boundary condition is different to ours,  their result itself  is not in conflict with ours.}

 The only background we know that allows arbitrary temperature at low temperature is the thermal AdS background. But,  this background does {\it not} have any charge.
How can we interpret this situation? Where the charges can go as temperature cools down? There are two options to interpret this:
\begin{enumerate}
  \item  If we keep the charge fixed, there is no confinement phase transition as temperature goes down.
  \item We take thermal AdS as a reference background
for the low temperature and take care of the charge conservation by IR-brane.
\end{enumerate}
Let's consider the former possibility: When we do not introduce the hard wall this is what happens as we discussed in section 3. However, in the presence of hard wall or confinement, it is physically implausible for a few reasons. First, in the absence of charge we know there is a confinement phase transition \cite{herzog}. Now, suppose phase transition disappear by adding baryons as it says.  Then, this phenomena is independent of how much charge is there, so even in the extreme low density disappearance of HPT is implied in this case.
How just a few charges can change the phase diagram of a system discontinuously?
Secondly, the confinement is determined by the dynamics of gluon degree of freedom which are ${\cal O}(N_c^2)$, and we do not expect it can be abolished by  introducing quark degree of freedom which is just ${\cal O}(N_c)$.
Therefore in the presence of confining
set-up in low temperature, we do not take the first possibility.
%\footnote{
%If one compactify one spacial direction, then one can %find an Euclidean solution that can compete with AdS-RN %backfround. However this would describe the 2+1 %dimensional physics not 3+1. We will describe this in the %next section.}

The second possibility is the only one left and it corresponds to storing the charges outside the repulsive gravitational  core. In this paper, for the convenience of treatment,
we  assume that all charges are at the IR brane located at $r_m$. With this set up, we can now calculate the phase diagram.

\begin{figure}[!ht]
  \begin{center}
\subfigure[]
{\includegraphics[angle=0,
  width=0.45\textwidth]{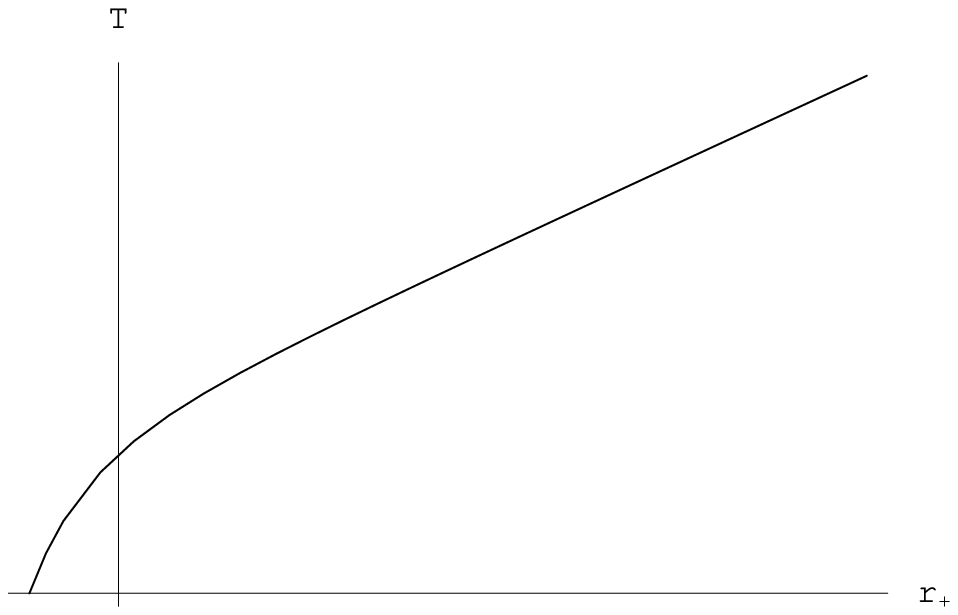} \label{fig:Tr}}
\subfigure[] {\includegraphics[angle=0,
width=0.45\textwidth]{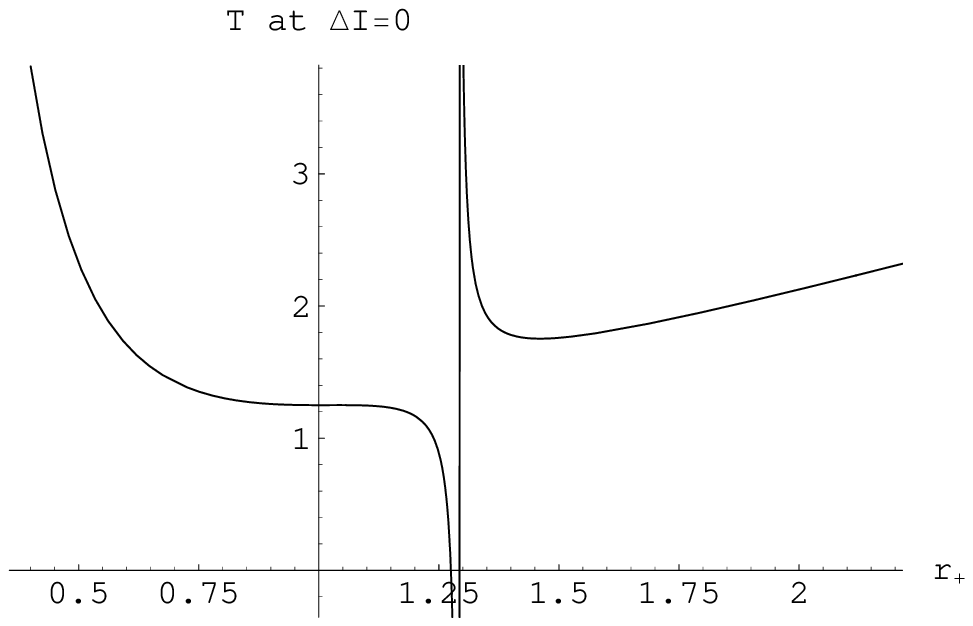} \label{fig:TrBd}}
  \caption{\label{fig:temperature}  Temperature as function of $r_+$. (a) fixed q, (b) $T$ v.s $r_+$ at the phase boundary.
  }
  \end{center}
  \end{figure}

\begin{figure}[!ht]
  \begin{center}
\subfigure[]
{\includegraphics[angle=0,
  width=0.45\textwidth]{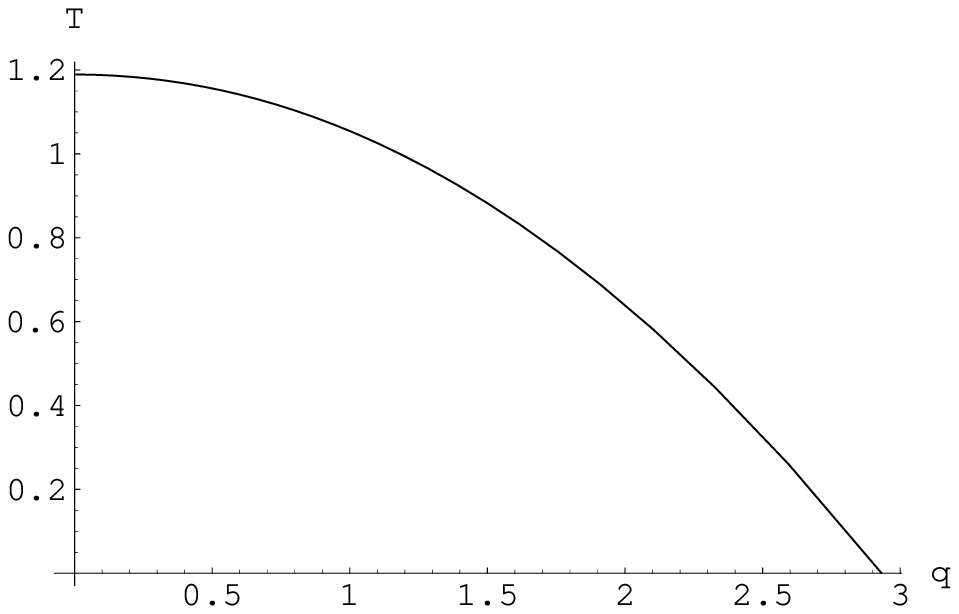} \label{fig:Tq}}
\subfigure[] {\includegraphics[angle=0,
 width=0.45\textwidth]{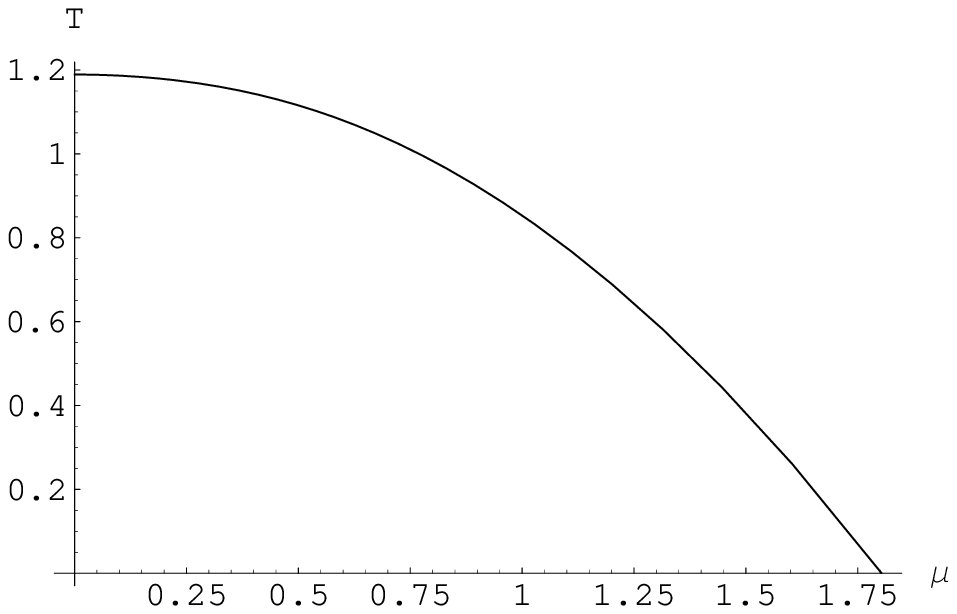} \label{fig:Tvsmu}}
  \caption{\label{fig:TvsQ} Phase diagrams  for the fixed charge (a) and  for the fixed chemical potential (b).
In  $q$-$T$ space,  two  ends points of phase boundary  are given by $(q,T)=(0,2^{1/4}r_m/\pi R^2)$ and $(2.9314r_m^3/R,0)$.
Below (over) the boundary is the (de-)confinement region.}
  \end{center}
  \end{figure}

The value of AdS action with charge at IR brane is given by
\ba
I_{AdS+Q}&=&\frac{\beta V_3}{2\kappa^2 R^3}\(-m + \frac{2(L^4-r_m^4)}{R^2} + \frac{2\kappa^2}{g^2}\frac{Q^2}{r_m^2}\)  , \label{fixedQAdS}\ea
where $m$ comes from the condition the temperature of AdS was tuned by  \eq{match}. Now the difference of the actions of two phases is
\ba
\Delta I&=&\frac{\beta V_3}{2\kappa^2 R^3}\[m - \frac{2(r_+^4-r_m^4)}{R^2}+\frac{q^2}{r_+^2} + \frac{2\kappa^2Q^2}{g^2}\(\frac{1}{r_+^2}-\frac{1}{r_m^2}\)\],\;\;  {\rm for} \;\; r_+>r_m \\
 &=&\frac{\beta V_3}{2\kappa^2 R^3}\[ m+ \frac{q^2}{r_m^2}  \]>0,\;\; {\rm for} \;\;  r_+<r_m
.\ea
The phase boundary is given by  $\Delta S=0$.
It is helpful to figure the temperature as a function of the horizon radius $r_+$.
See figure \ref{fig:temperature}.
Notice that unlike the round boundary case, temperature is
 monotonically increasing function of $r_+$ for all fixed value of $q$ starting from $r_+=q^{1/3}/2^{1/6}$ where $T=0$ (See figure \ref{fig:Tr}). However it is monotonically (rapidly) decreasing function of $r_+$
 for $2^{1/4}<r_+/r_m<1.27517$ which is
the relevant region for the phase boundary. ( See figure \ref{fig:TrBd}.)
The phase boundary  is parametrically given by
\be
{\tilde q}={\tilde r_+}\sqrt{\frac{{\tilde r_+}^4-2}{5-3{\tilde r_+}^2}},\quad
 {\tilde T}={\tilde r_+}\(1-\half \frac{{\tilde r_+}^4-2}{{\tilde r_+}^4(5-3{\tilde r_+}^2)}\),
\ee
with ${\tilde r_+}=r_+/r_m$, ${\tilde q}=qR/r^3_m$ and ${\tilde T}=\pi TR^2/r_m $. See figure \ref{fig:Tq}.

Notice that the phase diagram closes at finite density unlike the case without gravity back reaction\cite{HT,KLNPS}, where phase diagram is open.
The main reason of this change is attributable to
the change of the definition of the temperature by the second term of eq. \eq{temp}.

%\section{Density dependence of Physical Quantities}
%\section{ Constituent quark mass}
%The mass of the constituent quark mass in the D3/D7 %system is defined as the difference of black hole branch %and the Minkowski branch free energy.
%This make sense since in small temperature or large quark %mass limit, the difference is precisely equal to

\subsection{Fixed chemical potential}

When we fix chemical potential, charge itself is not
fixed. Only average charge is fixed.
So one might choose, as  the  low temperature phase, the thermal AdS without charge.
The phase diagram of resulting system  is worked out
in the appendix.
The issue here is that as a consequence of the choice, there would  be  no chemical potential dependence of the system in the low temperature phase. One might  argue that  $\mu$ independence is analogue of the  temperature independence of the low temperature phase of large $N_c$ limit.  For the round boundary, this choice was actually taken in \cite{EJM}. In \cite{parnachev} this consequence is used to support the chemical potential independence in chiraly asymmetric case.

However, comparing confinement  phase relative deconfinement phases,
thermodynamics functions of baryons are not so suppressed compared with that of glueballs.
The gluons are adjoint so  degree of freedom in confined phase relative to that in deconfined phase is ${\cal O} (1/N_c^2)$, while for baryon case,
it is suppressed by  ${\cal O}(1/N_c)$. So it is not clear whether baryon density independence can be implied by
the temperature independence, especially when we consider the gravity back reaction.
Also, if there is no charge in the system at all,
electric potential is globally constant and
if we calculate the average charge, it is zero.
\be \langle Q\rangle=-\(\frac{\d \Omega}{\d \mu}\)_{\!\beta}=0.\ee
It vanishes because there is no chemical potential dependence of the system.  It is not sensible to have
zero charge density for any chemical potential, since the latter case will have charge dependence as we will show shortly.
 Furthermore, this means the relation with fixed charge case by the Legendre transformation would be lost.
Therefore we should not  take the  pure AdS space describe  the low temperature phase. In this paper, we take the view that we should take the Legendre transform of the fixed charge system, which   discussed in the previous subsection. Then, the only change is replacing
$q \to aR r_+^2\mu$, whose jacobian  does not involve any singularity over the $r_+$ region of the phase boundary.
The phase diagram has the same topology as one can see in figure \ref{fig:Tvsmu}.
So we do not repeat the analysis here.

The point of interest is that the thermodynamic potential
defined by the Legendre transformation of the previous sections have non-trivial chemical potential dependence,
simply because the fixed charge case has non-trivial charge dependence. This is the answer to the issue  we raised in the beginning of this section.

Finally one may want to compare the phase diagram in this section with those in the appendix, where we analyzed the
choice  taking  the AdS space with no charge as the low temperature background.

\section{Discussion}

In this paper, we considered the back reaction of
the metric to the baryon charge for the case where
$AdS_5$ bulk is completely filled with flavor brane.
For the high temperature phase, the unique gravity
solution coupled with $U(1)$ electric potential is the charged AdS black hole with flat boundary. We also  addressed the issue whether we have a chemical potential dependence in low temperature   confining phase and concluded there is non-trivial chemical potential dependence.  Phase diagram for hard wall model in the presence of baryon charge is worked out and it turns out that phase diagram closes at finite baryon density.
This is very different from the previous treatment without gravity back reaction.
We emphasize that the general consideration of gravity back reaction in top down models are non-trivial \cite{liu,kirsch} 
and the simplicity of present model is coming from the bulk-filling nature of the  flavor branes in the 
effective ads/qcd models in the bottom-up approach. 

We also want to point out that with the charge dependence of the metric, density dependence of physical quantities can be encoded in quadratic action of AdS/QCD without adding higher order $\alpha'$ correction, due to the metric change, while in brane embedding approach, one has to include the  higher order ${\cal O}(F^4)$ effect to see it \cite{KSZ2}. It is interesting to workout the density dependence of physical quantities (mass and couplings) in this model.

There is another issue in the baryon density in AdS/CFT
which is not treated in this paper.
IN reference, \cite{NSSY},  electric  charges
were added  with boundary condition such that adding charges does not change smooth surface structure.  The phases allowed in this assumption is rich.
Shortly after the paper \cite{NSSY}, the authors of ref. \cite{KMMMT}, pointed out that electric  charges on flavor branes are  fundamental strings' end points. The force balancing condition is imposed and it was concluded that due to the spiky structures of the surface, branes should always touch the horizon, even at very low temperature.
  As a consequence, even a tiny amount of baryons should change the whole phase diagram which is reminiscent to the situation treated here as a puzzle.
We believe that this issue need to be discussed
more carefully, since baryon vertex staying outside the
horizon can change the  conclusion very easily.
Here out set up is such that brane is filling the bulk so that we can not  address this issue. It will be treated in a separate paper \cite{NSSY2}.

\bigskip
\appendix
\section{Phase diagram for grand canonical system with  thermal AdS as the low temperature phase.}

The difference of actions is
\ba
I_{RN}-I_{AdS}&=&\frac{\beta V_3}{2\kappa^2 R^3}\( - \(\frac{r_+^4}{R^2}+\frac{q^2}{r_+^2}\) +\frac{2r_m^4}{R^2}\),\;\; {\rm for} \;\;  r_+>r_m \\
 &=&\frac{\beta V_3}{2\kappa^2 R^3}\(  \( \frac{r_+^4}{R^2}+\frac{q^2}{r_+^2} \) -\frac{2q^2}{r_m^2}\),\;\; {\rm for} \;\;   r_+<r_m
.\ea

\begin{figure}[!ht]
  \begin{center}
  \subfigure[] {\includegraphics[angle=0,
  width=0.45\textwidth]{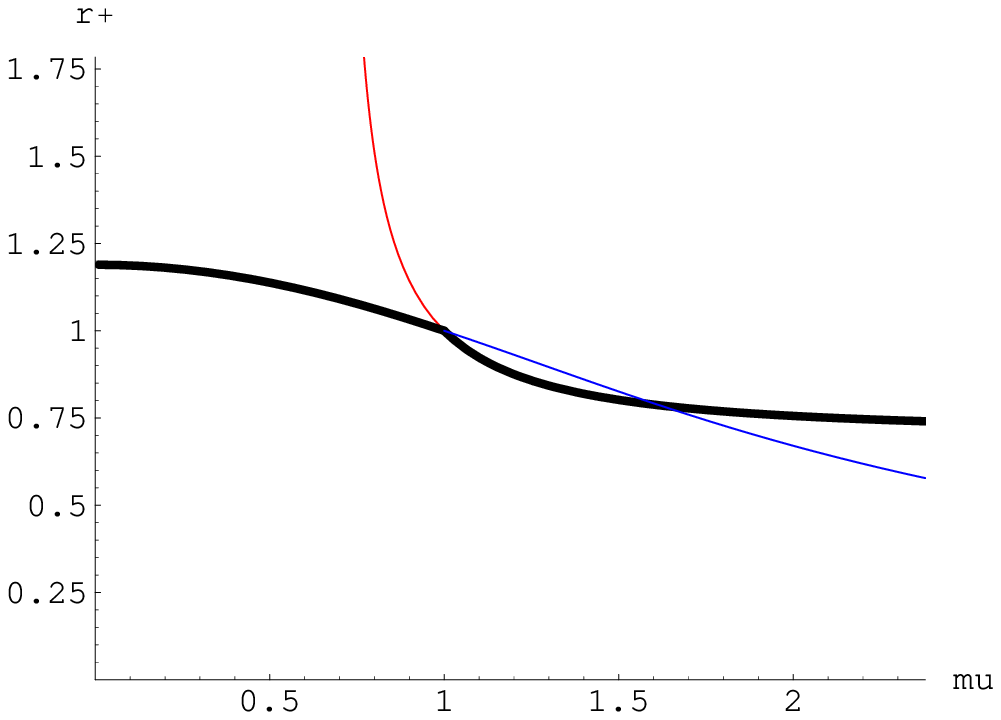} \label{fig:rplusmu}}
 \subfigure[] {\includegraphics[angle=0,
 width=0.45\textwidth]{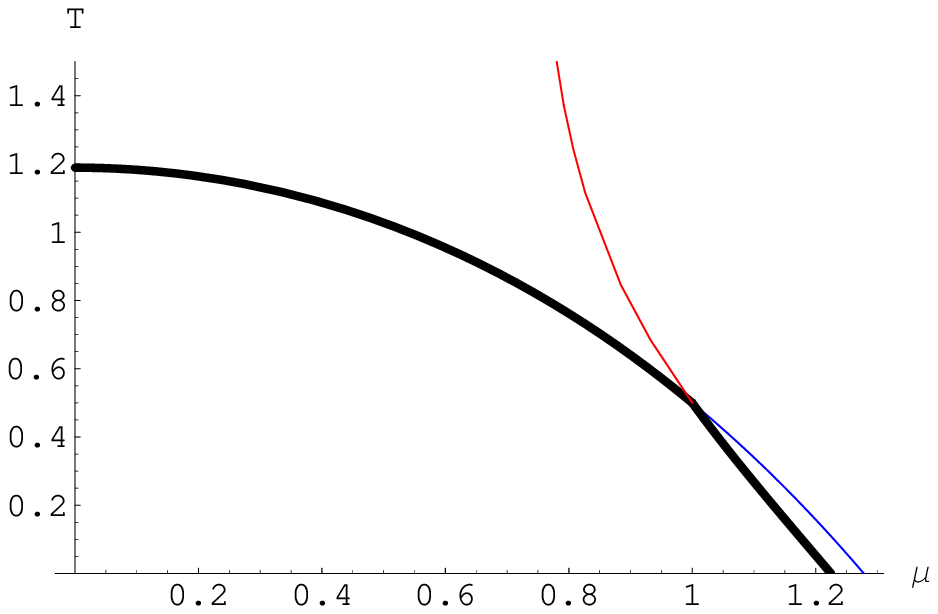} \label{fig:Tmu}}
  \caption{\label{fig:phase} Phase diagrams for the fixed chemical potential with  AdS without charge as the low temperature phase.
(a)$r_+ $ v.s $\mu$. In actual plot   $r$ means $r_+/r_m$ and $\mu$ means  $\mu aR/r_m$; (b) Temperature v.s chemical potential; Actual plot is $T aR/r_m$ v.s $\mu aR/r_m$. In each case, the thick line is the phase boundary and the connection point is $(T,\mu,r_+ )=( {r_m}/({2aR}),r_m/(aR),r_m)$ respectively.
Notice that phase diagram closes in $T$-$\mu$ diagram. }
  \end{center}
  \end{figure}

One can study the phase diagram by looking at the locus of the $\Delta S=0$:
\ba
  \( {r_+ }/{r_m}\)^4+    ({r_+/r_m})^2 (aR\mu/r_m)^2    &=& 2,\;\; {\rm for} \;\;   r_+>r_m , \;(\hbox{High T branch}),\\
  {r_m}^2/{r_+ }^2  +   r_m^2/(aR\mu)^2     &=& 2,\;\; {\rm for} \;\;  r_+<r_m,\;\; (\hbox{Low T branch})
.\ea
The phase diagram in $(\mu,r_+)$ is shown in figure. \ref{fig:rplusmu}.
Notice that there is a discontinuity in derivative at the connection point  $(\mu,r_+)=(r_m/(aR),r_m)$.

For the large chemical potential the phase boundary seems to saturate  asymptotically:
\be
r_+ \to \frac{r_m}{\sqrt{2}} \(1+  \frac{r_m^2}{4 a^2}\frac{1}{\mu^2}\), \ee
which is similar to the results in \cite{HT,KLNPS}, where
back reaction of the probe brane is not considered with identification $T=\frac{r_+}{\pi R^2}$. However, this similarity is just apparent one. In our case,
the temperature is given by
$T =\frac{r_+}{\pi R^2} -\frac{1}{2\pi}\frac{ (a\mu)^2}{r_+}$, and we need to consider the implication of the correction  given by the second term. For the positivity of temperature we need
$r_+\geq \frac{Ra\mu}{\sqrt{2}},$ which requires $r\geq \frac{\sqrt{3}}{2} r_m$ for the low temperature branch
 and $r\geq (2/3)^{1/4} r_m$ for high temperature branch.
One can easily work out the phase boundary in the parametric form:
\ba
{\tilde T}_H&=&{\tilde r_+}\(3/2-1/{\tilde r_+}^4\),\\
{\tilde \mu}_H&=& \sqrt{2-{\tilde r_+}^4}/{\tilde r_+},
\ea
for the high temperature branch and
\ba
{\tilde T}_L&=&{\tilde r_+}\frac{  {\tilde r_+}^2-3/4}{ {\tilde r_+}^2-1/2} ,\\
{\tilde \mu}_L&=& \frac{ {\tilde r_+}}{\sqrt{2 {\tilde r_+}^2-1}},
\ea
for the low temperature branch.
Here we used the  scaled variables, $ {\tilde r_+}=r_+/r_m$, $ {\tilde T}=TaR/r_m$ and $ {\tilde \mu}=\mu aR/r_m$.
By Plotting temperature v.s chemical potential, we see that the
phase diagram closes. See figure \ref{fig:Tmu}.
This difference of results can be easily understood: The chemical potential is related to the charge by $\mu=Q/r_+^2$ and charge is of order $N_f/N_c$, therefore in AdS/CFT with probe brane approach the range of the chemical potential is rather limited. However, our result indicates that, with large $N_f$ and large gravity back reaction, the phase diagram closes.

\acknowledgments
I would like to thank Keun-Young Kim,  Chanyong Park and Ismail Zahed  for  useful discussions.
This work was supported by KOSEF Grant R01-2004-000-10520-0 and
the SRC Program of the KOSEF through the Center for Quantum
Space-time of Sogang University with grant number R11-2005-021.

\end{document}